\documentclass[a4paper]{aa}
\usepackage{txfonts,graphicx,natbib}
\bibpunct{(}{)}{;}{a}{}{,}
\newcommand{\fo}{\ensuremath{f^\parallel}}
\newcommand{\fe}{\ensuremath{f^\perp}}

\newcommand{\FC}{\ensuremath{F_\mathrm{C}}}
\newcommand{\kk}{\ensuremath{k^\mathrm{(C)}}}

\newcommand{\pq}{\ensuremath{P_Q}}
\newcommand{\pu}{\ensuremath{P_U}}
\newcommand{\nnq}{\ensuremath{N_Q}}
\newcommand{\nnu}{\ensuremath{N_U}}

\begin{document}

\title{Broadband linear polarization of Jupiter Trojans}
       \author{
        S.~Bagnulo      \inst{1}
       \and
        I.~Belskaya     \inst{2}
       \and 
        A.~Stinson      \inst{1,3}
       \and
        A.~Christou     \inst{1}
       \and
        G.B.~Borisov    \inst{1,4}
              }
\institute{
           Armagh Observatory, College Hill, Armagh BT61 9DG, UK \\
           \email{sba@arm.ac.uk, ast@arm.ac.uk, aac@arm.ac.uk, gbb@arm.ac.uk}
           \and
           Institute of Astronomy, V.N. Karazin Kharkiv National University,
           35 Sumska str., 61022 Kharkiv, Ukraine.\\
           \email{irina@astron.kharkov.ua}
           \and
           Mullard Space Science Laboratory, University College London, 
           Holmbury St.\ Mary, Dorking RH5 6NT, UK
           \and
           Institute of Astronomy and National Astronomical Observatory,
           Bulgarian Academy of Sciences, 72, Tsarigradsko Chaussee Blvd., BG-1784 Sofia, Bulgaria
           }
\date{Received: 2015-07-03 / Accepted: 2015-09-29}
\abstract
{
Trojan asteroids orbit in the Lagrange points of the system
Sun-planet-asteroid. Their dynamical stability make their physical
properties important proxies for the early evolution of our solar
system.
}
{
To study their origin, we want to characterize the surfaces of
Jupiter Trojan asteroids and check possible similarities with
objects of the main belt and of the Kuiper Belt.
}
{ 
We have obtained high-accuracy broad-band linear polarization
measurements of six Jupiter Trojans of the L4 population and tried to
estimate the main features of their polarimetric behaviour. We have
compared the polarimetric properties of our targets among themselves,
and with those of other atmosphere-less bodies of our solar system.
}
{
Our sample show approximately homogeneous polarimetric behaviour,
although some distinct features are found between them. In general, the
polarimetric properties of Trojan asteroids are similar to those of
D- and P-type main-belt asteroids. No sign of coma activity is
detected in any of the observed objects.
}
{
An extended polarimetric survey may help to further investigate
the origin and the surface evolution of Jupiter Trojans.
}
\keywords{Polarization -- Scattering -- Minor planets, asteroids: general}

\maketitle
\section{Introduction}
Trojan asteroids are confined by solar and planetary gravity to orbiting
the Sun 60\degr\ ahead (L4 Lagrange point of the binary system
planet Sun) or behind (L5 Lagrange point) a planet's position along
its orbit \citep{murder99}. Stable trojans are supported by Mars, by Jupiter,
by Neptune, and by two Saturnian moons. Because of their dynamical
stability, they allow us to look at the earliest stages of the
formation of our solar system.  Saturn and Uranus do not have a stable
Trojan population because their orbits are perturbed on a short time
scale compared to the age of the solar system. Terrestrial planets may
support a population of Trojan asteroids, 
\citep[e.g.\ the Earth Trojan 2010 TK$_7$ discovered by][]{Conetal11},
but so far no stable population has been identified.

More than 6000 Trojans of Jupiter are known so far \citep{Emeretal15}.
In the framework of the Nice model of the formation of our solar
system, \citet{Morbetal05} predicted the capture of Jupiter
Trojans from the proto Kuiper belt. While these predictions were
invalidated by further simulations \citep[e.g.][]{NesMor12},
\citet{Nesetal13} investigated the possibility of the capture of Jupiter
Trojans from the Kuiper-belt region within the framework of the
so-called jumping-Jupiter scenario, and succeeded at reproducing
the observed distribution of the orbital elements of Jupiter Trojan
asteroids. The model by \citet{Nesetal13} supports the scenario in
which the majority of Trojans are captured from the trans-Neptunian
disk, while a small fraction of them may come from the outer asteroid
belt.

Unfortunately, direct spectral comparisons between the optical
properties of Jupiter Trojans with those of Kuiper-belt objects (or
trans-Neptunian objects, TNOs) show significant differences. TNOs have
a wide range of albedos that extend, in particular, to higher albedos,
while all known Jupiter Trojans have a low albedo and fairly
featureless spectrum, all belonging to `primitive' taxonomies,
principally C-, D-, and P-types of the Tholen \citep{Tholen84}
classification system \citep{Graetal12}. These types are the
most common ones in the outer part of the main belt.

\citet{Emeretal11} investigated the infrared properties of
Jupiter Trojans and report a bimodal distribution of their spectral
slopes. This bimodality is also seen in the albedos in the
  infrared \citep{Graetal12}, although it is not apparent in the
  optical albedo distribution. \citet{Emeretal11} interpret the slope
  bimodality as the observational evidence of at least two distinct
populations of objects within the Trojan clouds where the `less red'
group originated near Jupiter (i.e. either at Jupiter's radial distance
from the Sun or in the Main Asteroid Belt), while the `redder'
population originated significantly beyond Jupiter's orbit (where
similar `red' objects are prevalent). Therefore, at least the
near-IR spectroscopy observations of \citet{Emeretal11} are broadly
consistent with the widely accepted scenario suggested by
\citet{Morbetal05} and \citet{Nesetal13}, while the inconsistency in
the optical albedo and spectral properties could be naturally
explained by the fact that TNOs migrated to the Jupiter orbit have
been exposed to a different irradiation and thermal environment
\citep{Emeretal15}.

Polarimetric measurements are sensitive to the micro-structure and
composition of a scattering surface. In the case of the
atmosphere-less bodies of the solar system, the way that linear
polarization changes as a function of the phase angle (i.e., the angle
between the sun, the target, and the observer) may reveal information
about the properties of the topmost surface layers, such as the
complex refractive index, particle size, packing density, and
microscopic optical heterogeneity. Objects that display different
polarimetric behaviours must have different surface structures, so
that they probably have different evolution histories. Polarimetric
techniques have been applied to hundreds of asteroids
\citep[e.g.,][]{Beletal15}, as well as to a few Centaurs
\citep{Bagetal06,Beletal10} and TNOs
\citep{Boeetal04,Bagetal06,Bagetal08,Beletal10}. These works have
revealed that certain objects exhibit very distinct polarimetric
features. For instances, at very small phase angles, some TNOs and
Centaurs exhibit a very steep polarimetric curve that is not observed
in main-belt asteroids. This finding is evidence of substantial
differences in the surface micro-structure of these bodies compared to
other bodies in the inner part of the solar system.  It is therefore
very natural to explore whether optical polarimetry may help in
finding other similarities or differences among Jupiter Trojans, and
between Jupiter Trojans and other classes of solar system objects.

In this work, we carry out a pilot study intending to explore whether
polarimetry can bring additional constraints that help to understand
the origin and the composition of Jupiter Trojans better.
We present polarimetric observations of six objects belonging to
the L4 Jupiter Trojan population: (588)~Achilles, (1583)~Antilochus,
(3548)~Eurybates, (4543)~Phoinix, (6545)~1986~TR6, and
(21601)~1998~XO89. All our targets have sizes in the diameter range of
50--160\,km, and represent both spectral groups defined by
\citet{Emeretal11}.

From ground-based facilities, Jupiter Trojans may be observed up to a
maximum phase-angle of $\sim 12\degr$. Our observations cover the
range $7\degr-12\degr$ and are characterized by a ultra-high
signal-to-noise ratio (S/N) of $\sim 5000$, so their accuracy is not limited
by photon noise, but by instrumental polarization and other systematic
effects. Our observations are aimed at directly addressing the
question of how diverse the polarimetric properties of the L4
population of Trojans are and how they compare with the polarimetric
properties of other objects of the solar system. With our data we can
estimate the minimum of their polarization curves and
make a comparison with the behaviour of low-albedo main-belt
asteroids. Finally, by combining the polarimetric images, we can also
try to detect coma activity (if any) with great precision.

\section{Observations and results}
\begin{table*}
\caption{\label{Tab_Observations}
Polarimetry and photometry of six Jupiter Trojans asteroids in the
special $R$ FORS filter. $\pq$ and $\pu$ are the reduced Stokes
parameters measured in a reference system such that $\pq$ is the flux
perpendicular to the plane Sun-Object-Earth (the scattering plane)
minus the flux parallel to that plane, divided by the sum of the two
fluxes. Null parameters \nnq\ and \nnu\ are expected to be zero within
error bars. $m_R$ is the observed magnitude in the $R$ filter, and $R
(1,1,\alpha)$ is the magnitude as if the object was observed at
geocentric and heliocentric distances = 1\,au at phase angle
$\alpha$. Photometric error bars are estimated {\it a priori} = 0.05.
}
\begin{center}
\begin{tabular}{ccrlrr@{\,$\pm$\,}lrr@{\,$\pm$\,}lrrr}
\hline \hline
                                 & 
                                 & 
\multicolumn{1}{c}{}             & 
\multicolumn{1}{l}{}             & 
\multicolumn{1}{c}{Phase}        & 
\multicolumn{2}{c}{}             & 
\multicolumn{1}{c}{}             & 
\multicolumn{2}{c}{}             & 
\multicolumn{1}{c}{}             & 
\multicolumn{1}{c}{}             & 
\multicolumn{1}{c}{}            \\ 
\multicolumn{1}{c}{}             & 
\multicolumn{1}{c}{Time}         & 
Exp                              & 
\multicolumn{1}{c}{}             & 
\multicolumn{1}{c}{angle $\alpha$}&         
\multicolumn{2}{c}{\pq}          & 
\multicolumn{1}{c}{\nnq}         & 
\multicolumn{2}{c}{\pu}          & 
\multicolumn{1}{c}{\nnu}         & 
\multicolumn{1}{c}{}             & 
\multicolumn{1}{c}{}            \\ 
\multicolumn{1}{c}{Date}         & 
\multicolumn{1}{c}{(UT)}         & 
(sec)                            & 
\multicolumn{1}{l}{OBJECT}       & 
\multicolumn{1}{c}{(\degr)}      & 
\multicolumn{2}{c}{(\%)}         & 
\multicolumn{1}{c}{(\%)}         & 
\multicolumn{2}{c}{(\%)}         & 
\multicolumn{1}{c}{(\%)}         & 
\multicolumn{1}{c}{\ \ \ $m_R$}        & 
\multicolumn{1}{c}{$R(1,1,\alpha)$}      \\ 
\hline
            &       &      &            &      & \multicolumn{6}{c}{}                        &           &           \\
 2013 04 12 & 04:40 &  400 & 588        &  9.31& $-$1.07& 0.02&$-$0.02& $ $0.01& 0.02&$-$0.04&$    14.92$&$     8.39$\\      
 2013 04 18 & 01:16 &   96 & Achilles   & 10.03& $-$1.07& 0.04&$ $0.00& $-$0.07& 0.04&$-$0.06&$    15.00$&$     8.43$\\      
 2013 05 26 & 01:33 &  680 & (1906 TG)  & 11.93& $-$0.98& 0.03&$-$0.02& $-$0.01& 0.03&$ $0.01&$\le 15.59$&$\le  8.62$\\[2mm] 
 2013 04 11 & 02:26 &  560 & 1583       &  9.15& $-$1.22& 0.02&$ $0.02& $-$0.01& 0.02&$-$0.01&$    15.74$&$     8.87$\\      
 2013 04 18 & 04:13 &  480 & Antilochus &  9.75& $-$1.23& 0.03&$-$0.01&    0.01& 0.03&$ $0.02&$    15.89$&$     8.98$\\      
 2013 05 13 & 00:52 &  400 & (1950 SA)  & 11.07& $-$1.25& 0.03&$ $0.00&    0.00& 0.03&$ $0.02&$    15.96$&$     8.90$\\[2mm] 
 2013 04 12 & 03:31 & 1280 & 3548       &  7.35& $-$1.18& 0.03&$-$0.02& $-$0.04& 0.03&$ $0.03&$    16.73$&$     9.93$\\      
 2013 04 18 & 03:41 & 1450 & Eurybates  &  8.21& $-$1.25& 0.03&$-$0.05& $-$0.04& 0.03&$ $0.03&$    16.78$&$     9.95$\\      
 2013 04 19 & 04:31 & 1420 & (1973 SO)  &  8.35& $-$1.31& 0.03&$-$0.03&    0.02& 0.03&$-$0.04&$    16.72$&$     9.88$\\      
 2013 06 01 & 01:30 & 1760 &            & 11.18& $-$1.28& 0.04&   0.00&    0.03& 0.04&$-$0.08&$\le 17.29$&$\le 10.16$\\[2mm] 
 2013 04 11 & 03:00 & 1440 & 4543       &  7.32& $-$0.91& 0.03&$-$0.02& $ $0.00& 0.03&$-$0.02&$\le 16.71$&$\le  9.82$\\      
 2013 04 19 & 01:24 & 1660 & Phoinix    &  8.40& $-$0.91& 0.03&$ $0.02& $-$0.02& 0.03&$ $0.01&$    16.78$&$     9.85$\\      
 2013 06 04 & 01:56 & 1920 & (1989 CQ1) & 10.96& $-$0.97& 0.03&$-$0.02& $ $0.04& 0.03&$-$0.02&$    17.39$&$    10.14$\\[2mm] 
 2013 04 12 & 04:08 & 1080 & 6545       &  8.79& $-$1.20& 0.03&$ $0.03&    0.02& 0.03&$-$0.04&$    17.24$&$    10.43$\\      
 2013 04 25 & 02:58 & 2400 & (1986 TR6) & 10.13& $-$1.04& 0.09&$ $0.07&    0.11& 0.10&$-$0.01&$\le 17.37$&$\le 10.49$\\      
 2013 06 05 & 01:25 & 2400 &            & 11.14& $-$1.25& 0.04&$ $0.07&    0.02& 0.04&$-$0.01&$    17.81$&$    10.64$\\[2mm] 
 2013 04 11 & 03:40 & 1360 & 21601      &  6.83& $-$1.17& 0.03&$ $0.02& $-$0.02& 0.03&$-$0.01&$    16.84$&$    10.14$\\      
 2013 04 19 & 05:10 & 1390 & (1998 X089)&  7.85& $-$1.18& 0.03&$ $0.11&    0.01& 0.03&$-$0.06&$    16.93$&$    10.20$\\      
 2013 05 26 & 02:27 & 1920 &            & 11.07& $-$1.13& 0.09&$ $0.10& $-$0.07& 0.09&$-$0.01&$\le 17.67$&$\le 10.72$\\      
 2013 06 05 & 02:20 & 1920 &            & 11.36& $-$1.19& 0.03&$-$0.05&    0.04& 0.03&   0.05&$    17.59$&$    10.58$\\[2mm] 
\hline
\end{tabular}
\end{center}

\end{table*}
Our observations were obtained with the FORS2 instrument
\citep{AppRup92,Appetal98} of the ESO VLT using the well-established
beam-swapping technique \citep[e.g.][]{Bagetal09}, setting the retarder
waveplate at 0\degr, 22.5\degr, \ldots,157.5\degr.  For each observing
series, the exposure time accumulated over all exposures varied from a
few minutes for (588)~Achilles up to 40 minutes for (6545)~1986~TR6.

\subsection{Instrument setting}
Jupiter Trojans are relatively bright targets for the VLT, therefore
the S/N may be limited by the number of photons that can be measured
with the instrument CCD without reaching saturation, rather than by
mirror size and shutter time. The telescope time requested to reach an
ultra-high S/N is in part determined by overheads for CCD readout. The
standard readout mode of the FORS CCD has a conversion factor from
e$^-$ to ADU of 1.25 and 2$\times$2 binning readout mode. Each pixel
size, after rebinning, corresponds to 0.25\arcsec. Therefore, for a
1\arcsec\ seeing, the $2^{16}-1$ maximum ADU counts set by the ADC
converter limits the S/N achievable with each frame to $\sim
1000-1400$ (neglecting background noise and taking into account that
the incoming radiation is split
into two beams).  To increase the efficiency, we requested the use of
a non-standard $1 \times 1$ readout mode for our observing run. This
way, pixel size was reduced to 0.125\arcsec, and with a conversion
factor of ADU to e$^-$ of 1.25, we could expect to reach a S/N per
frame of $\sim 2000-2800$.  We also requested special sky flat fields
obtained with the same readout mode. While flat-fielding is not a
necessary step for the polarimetry of bright objects, we found that it
improved the quality of our results because it reduces the noise
introduced by background subtraction. For consistency with previous
FORS measurements of Centaurs and TNO, our broadband linear
polarization measurements were obtained in the $R$ filter.

\subsection{Aperture polarimetry}\label{Sect_Aperture_Polarimetry}
\begin{figure}
\begin{center}
\scalebox{0.35}{
\includegraphics*[angle=270,trim={0.9cm 0.7cm 0.3cm 0.8cm},clip]{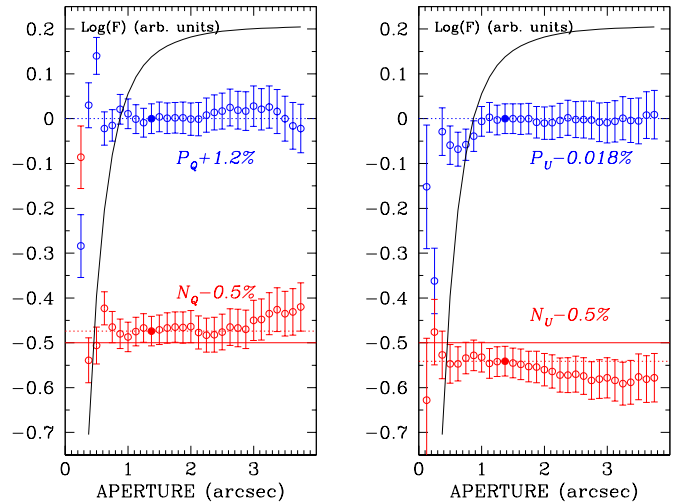}}
\caption{\label{Fig_Example_Apertures_QU} The principle of aperture
  polarimetry. The Figure is explained in the text (Sect.~\ref{Sect_Aperture_Polarimetry}
and \ref{Sect_Nulls}).}
\end{center}
\end{figure}
Fluxes were calculated for apertures up to a 30-pixel radius
(=3.75\arcsec) with one-pixel (=0.125\arcsec) increments. Sky
background was generally calculated in an annulus with inner and outer
radii of 28 and 58 pixels (i.e. 4.5\arcsec and 7.25\arcsec),
respectively. Imaging aperture polarimetry was performed as explained
in \citet{Bagetal11} by selecting the aperture at which the reduced
Stokes paremeters $\pq = Q/I$ and $\pu = U/I$ converge to a well-defined value. Polarimetric measurements are reported the perpendicular to the great circle passing
through the object and the Sun adopting as a
reference direction. This way, $\pq$ represents the flux
perpendicular to the plane Sun-object-Earth (the scattering plane)
minus the flux parallel to that plane, divided by the sum of these
fluxes. For symmetry reasons, $\pu$ values are always expected to be
zero, and inspecting their values allows us to perform an indirect
quality check of the \pq\ values.  This ``growth-curve'' method is
illustrated in Fig.~\ref{Fig_Example_Apertures_QU} for one individual
case, while Figs.~\ref{Fig_Apertures_PQ} and \ref{Fig_Apertures_PU}
(available in the appendix of the online version) show the same plot 
for all observed targets.

Figures~\ref{Fig_Example_Apertures_QU}, \ref{Fig_Apertures_PQ}, and
\ref{Fig_Apertures_PU} contain a lot of information, and it is
worthwhile commenting on them in detail.
In the left-hand panel of Fig.~\ref{Fig_Example_Apertures_QU}, the blue
empty circles show the \pq\ values measured as a function of the
aperture used for the flux measurement, with their error bars
calculated from photon noise and background subtraction
using Eqs.~(A3), (A4), and (A11) of \citet{Bagetal09}.
The \pq\ values are offset to the value adopted in
Table~\ref{Tab_Observations}. Ideally, for apertures slightly larger
than the seeing, \pq\ should converge to a well defined value, that
should be adopted as \pq\ measurement value in
Table~\ref{Tab_Observations}. 

Practically speaking,
Figs.~\ref{Fig_Example_Apertures_QU} and \ref{Fig_Apertures_PQ} 
clearly show that \pq \ sometimes\ depends on the
aperture in a complicated way, mainly due to the presence of
background objects that enter into the aperture where flux is measured
(see e.g.\ the case of Eurybates observed on June 1 in
Figs.~\ref{Fig_Apertures_PQ} and \ref{Fig_Apertures_PU}.) The values
reported in Table~\ref{Tab_Observations} were selected through visual
inspection of Figs.~\ref{Fig_Apertures_PQ},
as the value corresponding to the smallest aperture of a ``plateau" of
the growth curve rather than to its asymptotic value.
 
Lower in the figure, the empty red circles show the null parameters
offset to the value adopted in Table~\ref{Tab_Observations}, and
offset by $-0.5$\,\% for display purpose. The solid circle shows the
aperture adopted for the \pq\ measurement, and the corresponding
\nnq\ value is shown with a dotted line. In practice, the distances
between the solid line at $-0.5$\,\% and the empty circles correspond
to the null parameter values, and the distance between solid line and
the dotted line shows the \nnq\ value of Table~\ref{Tab_Observations}.
The physical significance of the null parameters is discussed in
Sect.~\ref{Sect_Nulls}.

The black solid line shows the logarithm of the total flux expressed
in arbitrary units. In this context it does not have any diagnostic
meaning, but demonstrates that polarimetric measurements converge at
lower aperture values than photometry and suggests that simple {\it
  aperture polarimetry} leads to results more robust than those of
{\it aperture photometry}.

The right-hand panels of Figs.~\ref{Fig_Example_Apertures_QU}
and~\ref{Fig_Apertures_PU} refer to \pu\ and \nnu and are organized in
exactly the same way as the left-hand panels of
Figs.~\ref{Fig_Example_Apertures_QU} and~\ref{Fig_Apertures_PQ},
respectively. For quality-check purposes, the aperture of \pu\ was
selected to be identical to that of \pq\ (see Sect.~\ref{Sect_PU}).

\subsubsection{Quality checks with the null parameters}\label{Sect_Nulls}
The polarimetric measurements presented here were obtained using
the so-called beam-swapping technique; i.e., Stokes parameters
are obtained as the difference between two observations obtained
at position angles of the retarder waveplate separated by 45\degr. 
This technique allows one to minimize spurious contributions due to the instrument. 
For instance, the reduced Stokes parameter \pq\ was obtained as
\begin{equation}
\frac{1}{2} \left[\pq(\phi=0\degr) + \pq(\phi=90\degr)\right]
,\end{equation}
where
\begin{equation}
\pq(\phi) =
\frac{1}{2}
\left[\left(\frac{\fo - \fe}{\fo + \fe}\right)_{\phi} - 
      \left(\frac{\fo - \fe}{\fo + \fe}\right)_{\phi+45\degr}\right]
\label{Eq_Basic}
,\end{equation}
where $\phi$ is the position angle of the retarder waveplate, and
\fo\ (\fe) is the flux measured in the parallel (perpendicular)
beam of the retarder waveplate.

The null parameter \nnq\ (\nnu) is defined as the difference between
the \pq\ values obtained from distinct pairs of observations:
\begin{equation}
\frac{1}{2} \left[\pq(\phi=0\degr) - \pq(\phi=90\degr)\right] \;.
\end{equation}
\citet{Bagetal09} have shown that, in the
ideal case, the results of repeated measurements of the null
parameters are expected to be scattered about zero according to a
Gaussian distribution with the same $\sigma$ as the \pq\ (\pu) error
bar. (Of course, we do not expect a Gaussian distribution for the
\nnq\ values measured on the same frames but with different apertures,
since these are not independent measurements.) The consistency of the
null parameters with zero within the \pq\ error bars are therefore an
indirect form of quality check.  For instance, an \nnq\ value
inconsistent with zero could be due to the presence of a cosmic ray,
or a background object or reflection in the aperture for some
positions of the retarder waveplate. These events would also affect
the \pq\ measurement, therefore one has to be wary of
\pq\ measurements that have high \nnq\ values.
Figures~\ref{Fig_Example_Apertures_QU}, \ref{Fig_Apertures_PQ}, and
\ref{Fig_Apertures_PU} show that the null parameters
are scattered about zero well within 2\,$\sigma$.

\begin{figure}[ht]
\begin{center}
\rotatebox{270}{
\scalebox{0.35}{
\includegraphics*[trim={0.9cm 1.6cm 0.3cm 0.8cm},clip]{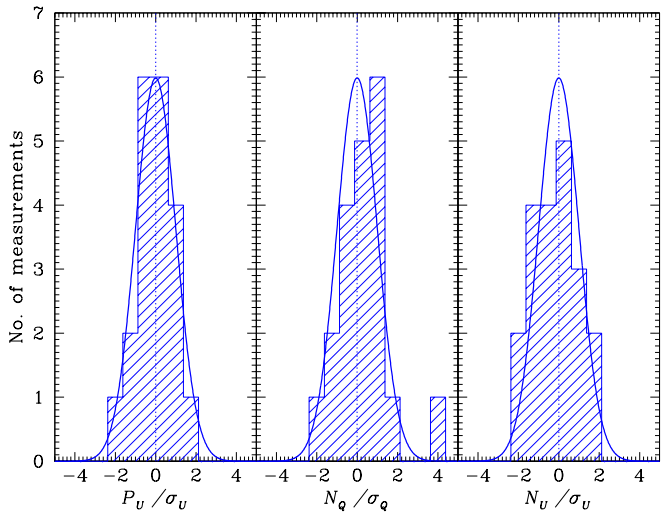}}}
\caption{\label{Fig_Histograms} Distribution of the \pu, \nnq\ and 
\nnu\ parameters normalised by the error bars, compared to a Gaussian
with $\sigma=1$.}
\end{center}
\end{figure}
\subsubsection{Quality checks with \pu}\label{Sect_PU}
If the target is macroscopically symmetric about the scattering plane,
\pu\ is expected to be zero. An individual \pu\ measurement that
significantly deviates from zero means either that there is a problem
with the measurement (similar to what is discussed for the null
parameter) or that the object is not symmetric about the plane
identified by the object, the Sun, and the observer. When considering
a large sample, all \pu\ measurements of different objects should be
scattered around zero. The null parameters should be scattered around
zero, and the null parameters normalized by their error bar should be
fit by a Gaussian with $\sigma=1$ centred on zero.

Inspection of the distribution of \pu\ parameters initially showed a
systematic offset by $\sim -1\,\sigma$. Another way to see exactly the
same effect is to calculate the average polarization position angle
measured from the perpendicular to the scattering plane: we found
90.4\degr instead of 90.0\degr. This 0.4\degr\ rotation offset can be
easily explained by an imperfect alignment of the polarimetric optics
and by an imperfect estimate of the chromatism of the retarder
waveplate.  To compensate for the waveplate chromatism in the R Bessel
filter, we had originally adopted the rotation suggested by the FORS
user manual of $-1.2\degr$. After inspecting the \pu\ values, we
instead decided to adopt a rotation by $-0.8$\degr.
Figure~\ref{Fig_Histograms} shows the histograms of the \pu, \nnq, and
\nnu\ values normalized to their error bars. The marginal deviations
from the expected Gaussian distribution do not look systematic and may
only be ascribed to the sample still being relatively small
statistically.

\subsection{Aperture photometry}\label{Sect_Aperture_Photometry}
The importance of acquiring simultaneous photometry and polarimetry
has probably been underestimated in the past. Modelling attempts need both pieces of
information, which are only available for a handful of
asteroids. However, at least with certain instrument configurations,
photometry may be a by-product of polarimetric measurements.  In the case
of the FORS instrument, an acquisition image is always obtained prior
to inserting the polarimetric optics. This can be used to
estimate the absolute brightness of the target, if the observing night
is photometric. (In fact, even if this is not the case, one could in
principle observe the same field again during a photometric night and
calibrate the previous observations). Therefore we performed aperture
photometry from our acquisition images, and then we calculated
\[
R(r=1\,{\rm au}, \Delta=1\,{\rm au},\alpha) = m_R - 5\,{\rm Log}_{10}\left(r\,\Delta\right)
\]
where $r$ and $\Delta$ are the heliocentric and geocentric distances,
respectively, and $m_R$ is obtained from the instrument magnitude $m_R^{\rm (instr)}$ using
\[
m_R = m_R^{\rm (instr)} - k_R X - k_{VR}\, (V-R) X + {ZP}_R
\]
where $ZP_R$, $k_R$ and $k_{VR}$ are the zero point and the extinction
coefficient in the $R$ filter and the $(V-R)$ colour index tabulated
in the FORS2 QC1 database, respectively, and $X$ is the
airmass. Aperture photometry can also be performed on the images
obtained with the polarimetric optics in, if these are
calibrated. From a comparison between photometry obtained from the
acquisition images and photometry obtained from the polarimetric
images (obtained adding \fo\ and \fe), we estimated that zero points
of the frames obtained in polarimetric mode with the $R$ special
filter can be obtained by subtracting 0.31 from the zero points
obtained in imaging mode.

Both $ZP_R$ and $k_{VR}$ are night dependent (their values are
$\sim 28.28$ and $0.01$, respectively). Based on the night-to-night
variations, we {\it \emph{a priori}} assigned an error of 0.05 and 0.0005 to
the zero point and to the colour term, respectively.  ESO classifies
each night with the symbols {\bf S}(table), {\bf U}(known), or {\bf
  N}on stable. Unfortunately, only three out of our 20 observing
series were obtained during stable nights. The reason is that to
maximize the chances that our observations would be performed during the
desired time windows, we set only loose constraints on
sky transparency. However, since we obtained several frames during an
extended period of time (typically 30--60\,min), it is still possible
to roughly evaluate the stability of the atmospheric conditions at the
time of our observations. We also note that the  Line of Sight Sky Absorption Monitor 
(LOSSAM, available online through the ESO web site) shows
that most of the observing nights were actually clear.

\begin{figure*}[ht]
\begin{center}
\scalebox{0.76}{
\includegraphics*[angle=270,trim={1.5cm 1.9cm 1.9cm 0.5cm},clip]{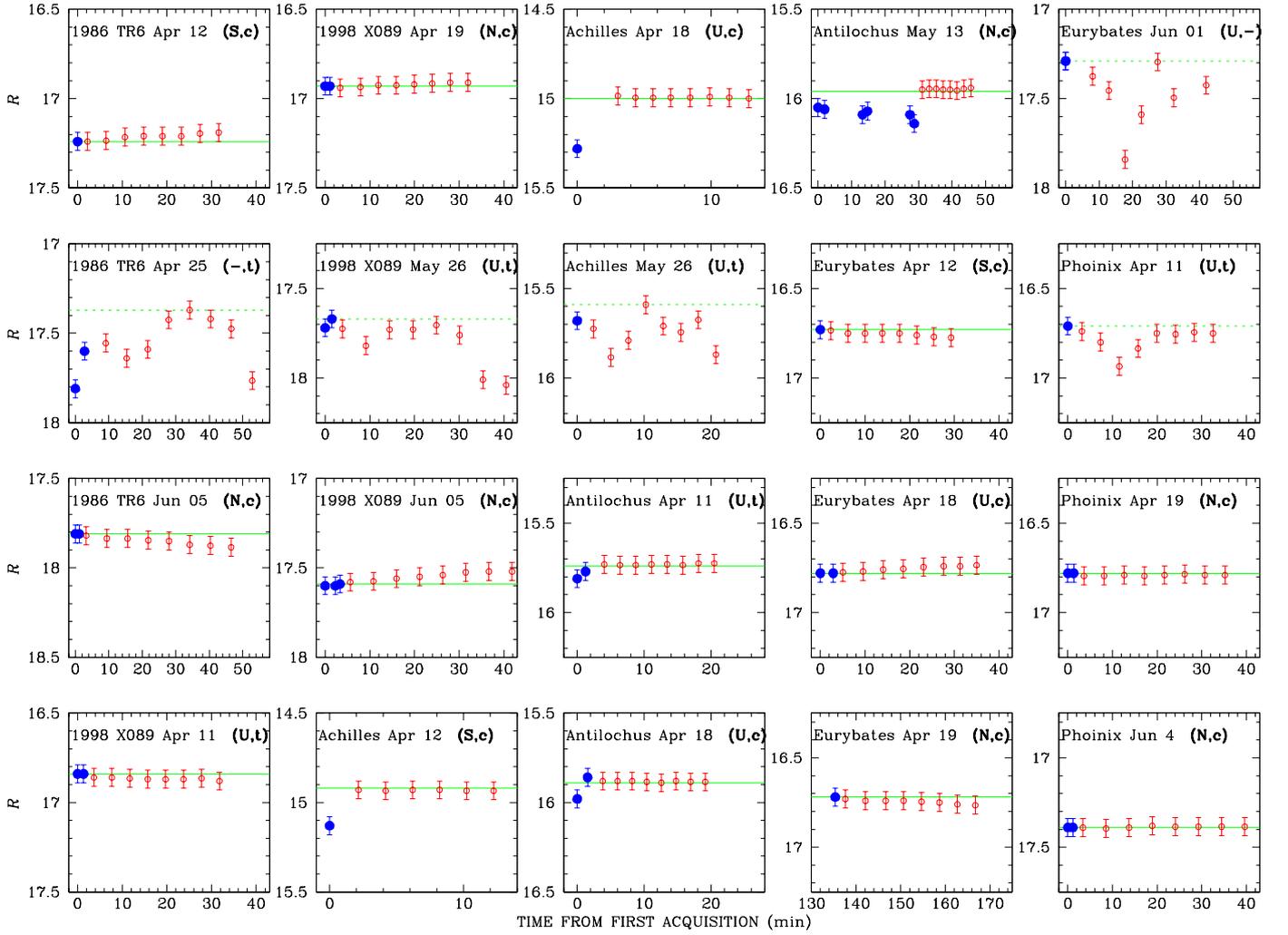}}
\caption{\label{Fig_Tab_Photometry} Photometric measurements.
The blue solid circles represent the photometry measured in the
acquisition images, and the red empty circles represent the
photometry measured from the polarimetric images. The green
lines show the final value adopted for the time series (dotted
green line represent an upper limit).
In each panel, between parenthesis we report the ESO QC1
classification of the night (U=unknown, N=non stable, S=stable)
followed by the sky conditions as we estimate after 
inspection of the LOSSAM plots available (c=clear,t=thin
to thick). The LOSSAM archive
is available online through the ESO website.
This Figure is discussed in Sect.~\ref{Sect_Aperture_Photometry}.
}
\end{center}
\end{figure*}

FORS acquisition images have a hard-coded $2 \times 2$ pixel readout
mode.  Aperture photometry was calculated on apertures up to 15 pixel,
and background was calculated in an annulus with inner radius of 20
and 30 pixels, respectively (corresponding to 5\arcsec\ and 7.5\arcsec).
The results of our photometric measurements are also reported in
Table~\ref{Tab_Observations}, and Fig.~\ref{Fig_Tab_Photometry} shows the magnitude measured
in each observing series.

\subsection{Searching for coma activity}\label{Sect_Coma}
\begin{figure}
\begin{center}
\scalebox{0.58}{
\includegraphics*[trim={2.3cm 8.7cm 1.0cm 9.0cm},clip]{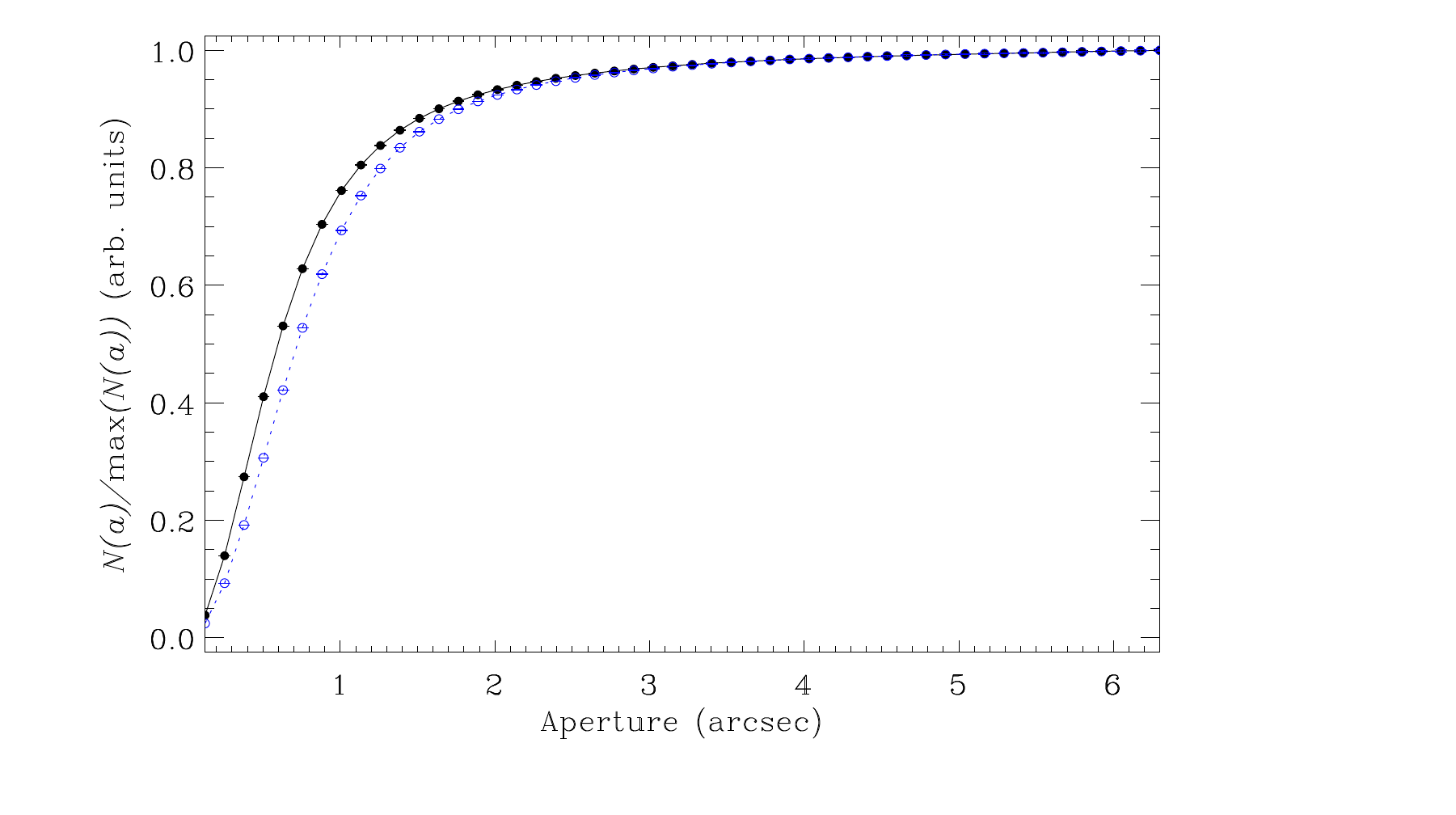}}\\
\scalebox{0.58}{
\includegraphics*[trim={2.3cm 8.7cm 1.0cm 9.0cm},clip]{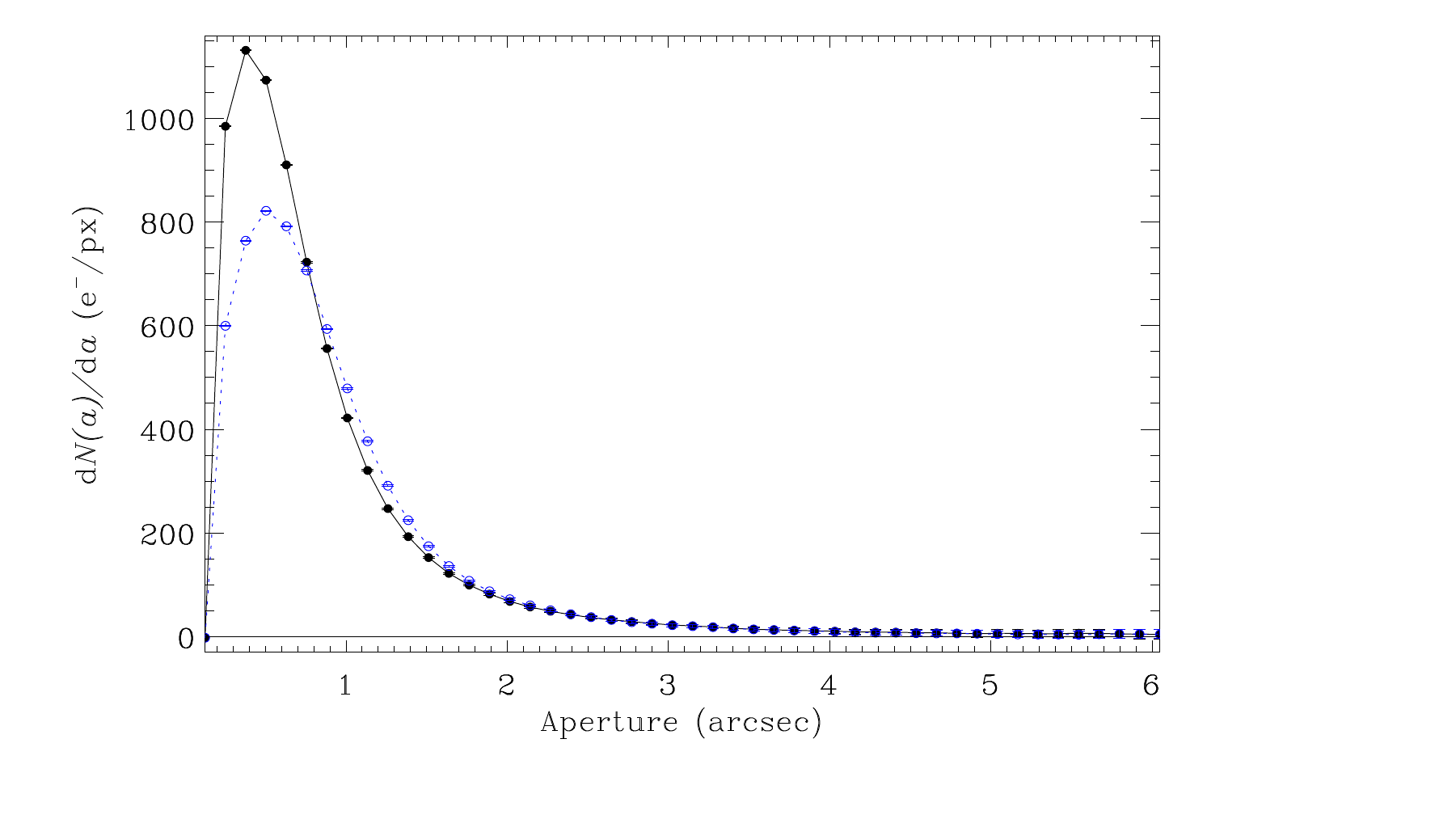}}
\caption{\label{Fig_Coma} Flux (top panel) and its derivative (bottom panel)
as a function of the aperture for asteroid (6545) 1986 TR6  observed on
2013-06-05 (black solid circles and solid lines) and for a background star of 
similar brightness (blue empty circules and dashed lines).
}
\end{center}
\end{figure}
After background subtraction, all polarimetric frames of each
observing series were coadded (combining together both images
split by the Wollaston prism and obtained at different positions of
the retarder waveplate). The resulting frames were analysed as
explained in Sect.~3.2 of \citet{Bagetal10} to check for the presence
of coma activity. Briefly, we assumed that the number of detected
electrons e$^-$ of the object per unit of time within a circular
aperture of radius $a$ is the sum of the contribution of the nucleus
plus the potential contribution of a coma, plus, possibly, a spurious
contribution due to imperfect background subtraction. To check for
the presence of a coma, it is probably sufficient to compare the
point-spread function (PSF) of the main target with those of the
background stars. However, if we are interested in a more quantitative
estimate (e.g. an upper limit), following \citet{AHeetal84}, we can
assume that the flux of a weak coma around the nucleus in a certain
wavelength band can be written as
\begin{equation}
\FC = A f  \left(\frac{\rho}{2 r \Delta}\right)^2  {F_\odot}\; ,
\label{Eq_Af}
\end{equation}
where $A$ is the bond albedo (unitless), $f$  the filling factor
(unitless), $r$  the heliocentric distance expressed in au, $\Delta$
 the geocentric distance and $\rho$  the projected
distance from the nucleus (corresponding to the aperture). Finally, $F_\odot$ is the solar flux at 1\,au, integrated in the same band
as \FC,   and convolved with the filter transmission curve. Following
the approach of \citet{TozLic02}, \citet{Bagetal10} have shown that
if the derivative of the flux with respect to the aperture converges
to a constant value \kk, then 
\begin{equation}
A f \rho = 1.234\,10^{19} \ 10^{0.4\,(m_\odot - {ZP}_m)} \
           r^2 \ \left(\frac{\Delta}{d_{\rm p}}\right) \ \kk
\label{Eq_Afrho}
,\end{equation}
where $m_\odot$ is the apparent magnitude of the Sun (i.e., at 1\,au)
in the considered filter, $ZP_m$ is the zero point in that filter for
the observing night, and $d_{\rm p}$  the CCD pixel scale in arcsec
(0.125\arcsec\ in our case). In Eq.~(\ref{Eq_Afrho}) $r$ and $\Delta$ are measured in au and $\kk$ in e$^-$ per pixel, and $A f \rho$ is
obtained in cm. We found that in all cases, $Af\rho$ is consistent
with zero within a typical error bar of $\sim 10$\,cm (see Fig.~\ref{Fig_Coma}).
We conclude that there is no evidence of any coma activity.

\section{Discussion}
\begin{figure}
\begin{center}
\scalebox{0.45}{
\includegraphics*[trim={0.7cm 5.5cm 0.0cm 2.5cm},clip]{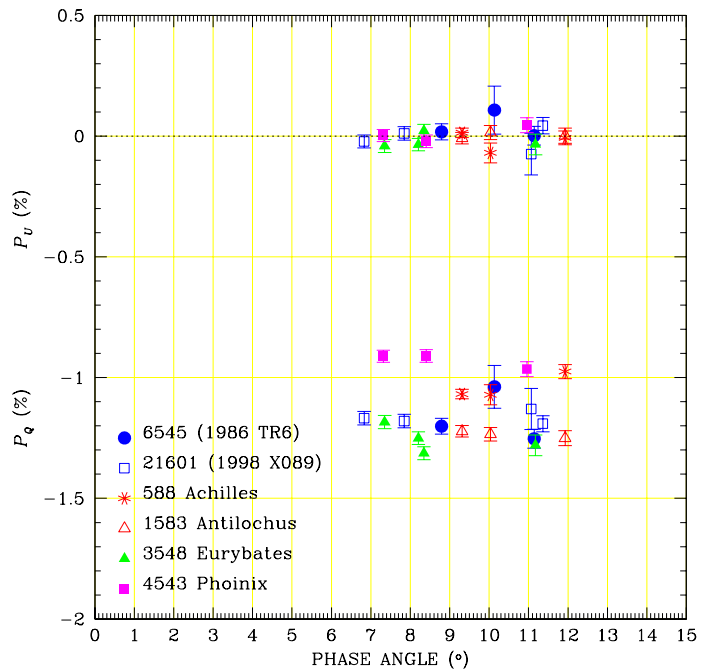}}
\caption{\label{Fig_Trojans} Polarimetric measurements of six Jupiter
Trojans asteroids.}
\end{center}
\end{figure}

All our polarimetric and photometric measurements are reported in
Table~\ref{Tab_Observations}.  In the following we first discuss the
differences found among our sample, searching for a correlation
between polarimetric properties and other characteristics of our
objects, then we consider Trojans as a homogeneous class to compare
with other atmosphere-less objects of the solar system.

Orbital constraints meant that all six Trojans asteroids were observed in
the negative branch, i.e. at those phase angles where we expect that
the polarization of the reflected light is parallel to the scattering
plane. We measured polarization values from $-1.3$\,\% to
$-0.9$\,\% in a phase-angle range $7-12\degr$.  The variations in
polarization within the observed phase-angle range are small for all
objects, yet, thanks to the high S/N of our observations, it is possible to distinguish some different behaviours.
Figure~\ref{Fig_Trojans} shows the results of our polarimetric
observations as a function of the phase angle.  

Several functions have been proposed to fit polarimetric measurements
versus phase angle. One of the most popular ones is the one proposed by
\citet{LumMui93}:
\begin{equation}
P(\alpha) = b \sin^{c_1}(\alpha)\,\cos^{c_2}\left(\frac{\alpha}{2}\right)\,
\sin(\alpha-\alpha_0)
\label{Eq_Lumme}
\end{equation}
where $b$ is a
parameter in the range $[0,1]$, $\alpha_0$ is the inversion
angle (typically $\la 30\degr$), and $c_1$ and $c_2$ are positive
constants.
Equation~(\ref{Eq_Lumme}) was used, for example, by \citet{Penetal05} for a
statistical study of the asteroids and comets. The number of our
data points per object is even smaller than the number of free parameters,
therefore it does not make sense to fit our data without making assumptions
(such as about the inversion angle of the polarimetric curves).
However, assuming that the minimum of the polarization is reached
in the phase-angle range 6\degr-12\degr \ (a typical range for low-albedo objects
would be 8\degr-10\degr),
even a simple visual inspection allow us to estimate the polarization minima
of the various objects and, in particular, to conclude that our sample
does not show homogeneous polarimetric behaviour.

The object (3548) Eurybates is the largest member of a dynamical family mainly
consisting of C-type objects \citep{Foretal07}. It has the deepest
minimum ($P_{\rm min}\sim-1.3$\,\%). All the remaining objects belong to the
D-type taxonomic class \citep{Graetal12}. The objects  (588) Achilles and (4543)
Phoinix exhibit a shallower polarization curve (i.e., lower absolute
values of the polarization) than the other four Trojans. Figure~\ref{Fig_Trojans}
also suggests that the minimum of the polarization curve of (588)
Achilles is reached at a phase-angle value that is lower than that of (4543)
Phoinix. Objects (1583) Antilochus, (6545) 1986 TR6, and (21601) 1998 X089
all seem to have similar polarimetric behaviour, with a minimum
$\sim -1.2$\,\%.

Before progressing in our analysis, it is important to discuss whether
the observed diversities are real. This question arises since our
photon-error bars are very small (a few units in $10^{-4}$), and
compared to them, instrumental or systematic errors may not be
negligible.  However, the relatively smooth behaviour with
phase angle and the good consistency with zero of both the \pu\ and
the null parameters suggest that our photon-noise error bars are
probably representative of the real error.  Exceptions to the smooth
behaviour are represented by the point at phase angle 10.1\degr of
asteroid (6545) 1986 TR6 (obtained on April 25 2013) and the point
at phase angle 8.4\degr\ of asteroid (3548) Eurybates (obtained on
April 19 2013). Figures~\ref{Fig_Apertures_PQ} and
\ref{Fig_Apertures_PU} show that, in the former case, polarimetric
measurements depend strongly on the aperture and fail to converge to
a well-defined value, probably due to the strong background, therefore
the observed discrepancy (still within the error bar) is due to a
larger error than what is typical in our dataset. The case of asteroid
(3548) Eurybates is more puzzling. There is nothing in
Fig.~\ref{Fig_Apertures_PQ} that suggests a problem with aperture
polarimetry in any of the observations, therefore one may hypothesize
that the abrupt change observed between the point at phase
8.1\degr\ and the point at phase angle 8.4\degr\ is due to asteroid
rotation.

The rotation periods of the observed Trojans range from 7.306\,h for
(588) Achilles to 38.866\,h (4543) Phoinix, and their ligtcurves
amplitudes are $\la 0.3$\,mag. While during an observing series we do
not expect short-term photometric variations caused by asteroid
rotation, it is possible that polarimetric data depend on the rotation
phase at which observations were obtained. 

Polarimetric behaviour may depend on the rotational phase of the
observations. Although rarely observed, one notable example is that of
asteroid (4) Vesta, with a rotational polarimetric amplitude of $\sim
0.03$\,\% \citep{WikNof15} to 0.1\,\% \citep{Lupetal88}. To test
whether our polarimetric data are rotationally modulated, we
calculated the rotation phase shift between observations of each
object and found, for instance, that a large shift (0.36) occurs
between the observations at phase angles 6.8\degr\ and 7.8\degr\ of
asteroid (21601) (1998 X089). However, the polarization values at
these phase angles are consistent with each other, and overall, the
polarization curve is relatively smooth.  By contrast, the rotation
phase shift between phase angles 8.2\degr and 8.4\degr\ of asteroid
(3548) Eurybates is only 0.1 of a rotation period.  These differences
may, therefore, not be due to rotation but instead to photon noise
fluctuations or to small changes in the (already small)
instrumental polarization. We conclude that there is no obvious
evidence of a polarimetric modulation introduced by asteroid rotation
in our data. On the other hand, our sample shows a polarimetric behaviour that is not
perfectly homogeneous, which must reflect some difference in their
surface structure and/or albedo.
\begin{figure}
\begin{center}
\scalebox{0.65}{
\includegraphics*[trim={2.0cm 2.0cm 0.0cm 19.0cm},clip]{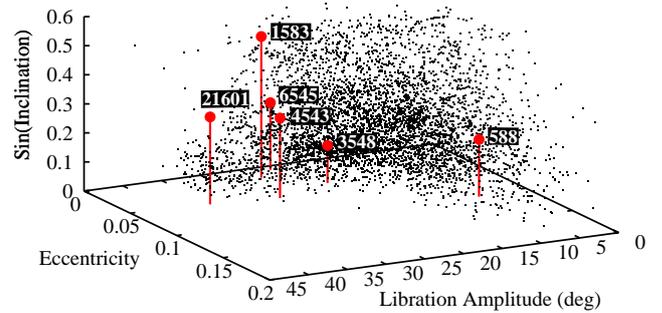}}
\caption{\label{Fig_Dynam} 
Proper orbital elements of 5020 numbered and multi-opposition Trojans
appearing in the database by Knezevic \& Milani
(last updated in June 2014; available at
{\tt http://hamilton.dm.unipi.it/$\sim$astdys2/propsynth/tro.syn}). 
The six Trojans
observed in this work are indicated by the red points and identified by
number.}
\end{center}
\end{figure}

\subsection{Searching for correlation with dynamical and surface properties}

\subsubsection{Proper orbital elements}
No strong correlations have been identified yet between the physical
and orbital properties of Trojans, although there appears to be a
bimodality in spectral slopes
\citep{Szaetal07,Roietal08}. In confirming a similar
bimodality within a sample of near-IR spectra of Trojans, \citet{Emeretal11}  point out a
possible weak correlation with inclination amongst their less-red
population. We have searched for trends between polarimetric behaviour
and orbital properties in our sample. Figure~\ref{Fig_Dynam} shows the
proper orbital elements\footnote{These are constants that parameterize
  the evolution of their {\it osculating} elements, the latter varying
  with time due to planetary perturbations (Milani, CeMDa, 1993)} of
Jupiter Trojans with the six objects observed in this work denoted by
red points and identified by their number.  All the objects, except
(588) Achilles, have low ($\lesssim 0.05$) proper eccentricity and
high ($> 15^{\circ}$) libration amplitude. We note that the two
objects with a shallow polarization curve -- (588)~Achilles and
(4543)~Phoinix -- also have the highest proper eccentricity in our
sample. However, since the eccentricity of (4543)~Phoinix ($0.059$) is
only marginally higher than those of (3548)~Eurybates and 21601 (1998
X089) ($0.044$ and $0.053$, respectively), and given the small size of
our sample, we do not attach any high statistical significance to this
observation. Finally, there appears to be no trend linking
polarization behaviour with inclination; both (588)~Achilles and
(4543)~Phoinix have inclinations within 1\,$\sigma$ of the mean for
the four other objects.

\begin{table*}
\caption{\label{Tab_Para} Some parameters of the observed Jupiter Trojans.}
\begin{small}
\begin{center} 
\begin{tabular}{rlrr@{$\pm$}lr@{$\pm$}lrrrr@{$\pm$}lr@{$\pm$}lcc}        
\hline\hline                 

      &     &                        
\multicolumn{1}{c}{}           &     
\multicolumn{4}{|c}{DIAMETER}   &     
\multicolumn{3}{|c|}{$H_V (\alpha=0\degr)$}&           
\multicolumn{6}{c}{ALBEDO ESTIMATES} \\ 
      &     &
\multicolumn{1}{c}{$P_{\rm min}$} &     
\multicolumn{2}{c}{WISE}       &      
\multicolumn{2}{c}{AKARI}      &      
\multicolumn{1}{c}{}       &      
\multicolumn{1}{c}{}      &      
\multicolumn{1}{c}{this}       &      
\multicolumn{2}{c}{} &   
\multicolumn{2}{c}{} &   
\multicolumn{1}{c}{$D_{\rm WISE}\ +$}  &   
\multicolumn{1}{c}{$D_{\rm AKARI}\ +$} \\  
Object& Type&
\multicolumn{1}{c}{(\%)}  &               
\multicolumn{2}{c}{(km)}  &               
\multicolumn{2}{c}{(km)}  &               
\multicolumn{1}{c}{WISE}      &               
\multicolumn{1}{c}{AKARI}      &               
\multicolumn{1}{c}{work}  &               
\multicolumn{2}{c}{WISE}     &    
\multicolumn{2}{c}{AKARI}     &    
\multicolumn{1}{c}{$H_{\rm this\ work}$}  &    
\multicolumn{1}{c}{$H_{\rm this\ work}$} \\    
\hline                       
\multicolumn{1}{c}{1}&
\multicolumn{1}{c}{2}&
\multicolumn{1}{c}{3}&
\multicolumn{2}{c}{4}&
\multicolumn{2}{c}{5}&
\multicolumn{1}{c}{6}&
\multicolumn{1}{c}{7}&
\multicolumn{1}{c}{8}&
\multicolumn{2}{c}{9}&
\multicolumn{2}{c}{10}&
\multicolumn{1}{c}{11}&
\multicolumn{1}{c}{12}\\
\hline
 588 & DU & $-1.10$ & 130.1&0.6 & 133.2&3.3 &   8.47 & 8.67 & 8.45 & 0.043&0.006& 0.035&0.002&  0.044 &0.042\\
1583 & D  & $-1.25$ & 108.8&0.5 & 111.7&3.9 &   8.60 & 8.60 & 8.97 & 0.054&0.004& 0.053&0.004&  0.039 &0.037\\
3548 & C  & $-1.35$ &  63.9&0.3 &  68.4&3.9 &   9.80 & 9.50 & 9.79 & 0.052&0.007& 0.060&0.007&  0.053 &0.046\\
4543 & D  & $-0.95$ &  63.8&0.4 &  69.5&2.2 &   9.70 & 9.70 &10.06 & 0.057&0.017& 0.049&0.003&  0.041 &0.035\\
6545 & D  & $-1.25$ &  51.0&0.6 &\multicolumn{2}{c}{}                                     
                                            &  10.00 &      &10.64 & 0.068&0.009&\multicolumn{2}{c}{}&0.038&\\
21601& D* & $-1.20$ &  54.9&0.4 &  56.1&1.9 &   9.90 & 9.40 &10.41 & 0.064&0.012& 0.100&0.007&  0.040 &0.039\\

\hline
\end{tabular}
\end{center}
\end{small}

\end{table*}

\subsubsection{Albedo}
It is well known that the minimum of the polarization is inversely
correlated to the albedo; i.e., the higher the absolute value of the
minimum, the lower the albedo \citep[e.g.][]{Zeletal77a,Celetal15}. In
fact, various authors have tried to calibrate a relationship
\begin{equation}
\label{Eq_Alpol}
\log p = C_1 \log P_{\rm min} + C_2
\end{equation}
to estimate the albedo $p$ from polarimetric observations. For instance,
\citet{LupMoh96} give $C_1 = -1.22$ and $C_2 = -0.92 $; \citet{Celetal15}
give $C_1 = -1.426 \pm 0.034$ and $C_2 = -0.917 \pm 0.006$.

However, it is known that Eq.~(\ref{Eq_Alpol}) is only an
approximation that does not necessarily produce accurate albedo
estimates \citep[e.g.][]{Celetal15}. In particular, a saturation effect may occur for the darkest
objects, which was discovered in the
laboratory for very dark surfaces \citep{Zeletal77b,Shketal92}: the
depth of negative polarization increases as the albedo decreases down
to $\sim 0.05$, but a further decrease of the albedo results in a
decrease in the absolute value of the polarization minimum. This
effect was observed for the very dark F-type asteroids by \citet{Beletal05}
\citep[see also][]{Celetal15}. In the case of the observed Trojans,
the albedo estimated from Eq.~(\ref{Eq_Alpol}) and from our polarimetric
minima are of the order of 0.08--0.12, which is inconsistent with what has been
found from independent estimates of the albedo. We conclude that the polarimetric
measurements of our Trojans are 
also in the regime of `saturation' similar to what
was observed for F-type asteroids.

In fact, the albedo estimates from the WISE \citep{Graetal12} and AKARI
\citep{Usuetal11} mid-IR surveys lead to contradictory
conclusions. For instance, according to AKARI data (Col.~10 of
Table~\ref{Tab_Para}), (588) Achilles and (4543) Phoinix (that show
the shallower polarization minima) are actually the darkest
objects. This finding is somehow contradicted by the WISE albedos
(Col.~9), according to which (588) Achilles would still be the darkest
object in our sample, but (4543) Phoinix would have an albedo higher
than that of (1583) Antilochous and (3548) Eurybates. Albedo estimates
strongly depend on the values of absolute magnitudes\footnote{The absolute
magnitude $H$ is the magnitude that would be measured in the $V$ filter 
if the asteroid was observed at geocentric and heliocentric distances =
1\,au and phase angle $\alpha=0$} adopted in the
surveys (see Cols. 6 and 7). It is therefore of some interest 
to recalculate them using our photometric
measurements in Table~\ref{Tab_Observations}.

To calculate the absolute magnitudes, we need to know magnitude-phase
dependences of our targets. \citet{Sheetal12} have shown that D-type
Trojans are characterized by a linear magnitude-phase dependence down
to small phase angles without the opposition effect, i.e., that a linear
fit gives a more precise estimate of the absolute magnitudes of the D-
and P-type Trojans compared to what can be estimated with the
so-called HG function \citep{Slyetal12}. For the D-type asteroids, we
therefore performed a linear extrapolation to zero phase angle
assuming a 0.04 mag/deg slope, which is typical of these objects.  For the
C-type (3548) Eurybates, we assumed a non-linear magnitude-phase
dependence similar to that of C-type asteroids \citep{BelShe00}. To
calculate the absolute magnitudes $H$ in the $V$ band, we adopted the
literature $V-R$ colours of these objects, when available, or assumed
$V-R=0.45$ \citep[see][]{Foretal07}. Our estimates of the absolute
magnitudes $H$ are shown in Col.~8 of Table~\ref{Tab_Para}. Although our
photometric measurements agree with the measurements of
Cols.~6 and 7, they exhibit a systematic negative offset, which may be
consistent with the findings by \citet{Praetal12} of a systematic bias
in the absolute magnitudes of asteroids given in the orbital
catalogues. Using our revised absolute magnitudes and diameters from
WISE and AKARI surveys, we calculated the albedos of our
objects again. Our new albedo estimates (Cols.\ 11 and 12 of
Table~\ref{Tab_Para}) are no longer as scattered as the original
estimates from \citet{Usuetal11} and \citet{Graetal12}, but actually
very similar for all five D-type Trojans.
\begin{figure}
\includegraphics*[scale=0.9,trim={0.0cm 0cm 0.0cm 0.5cm},clip]{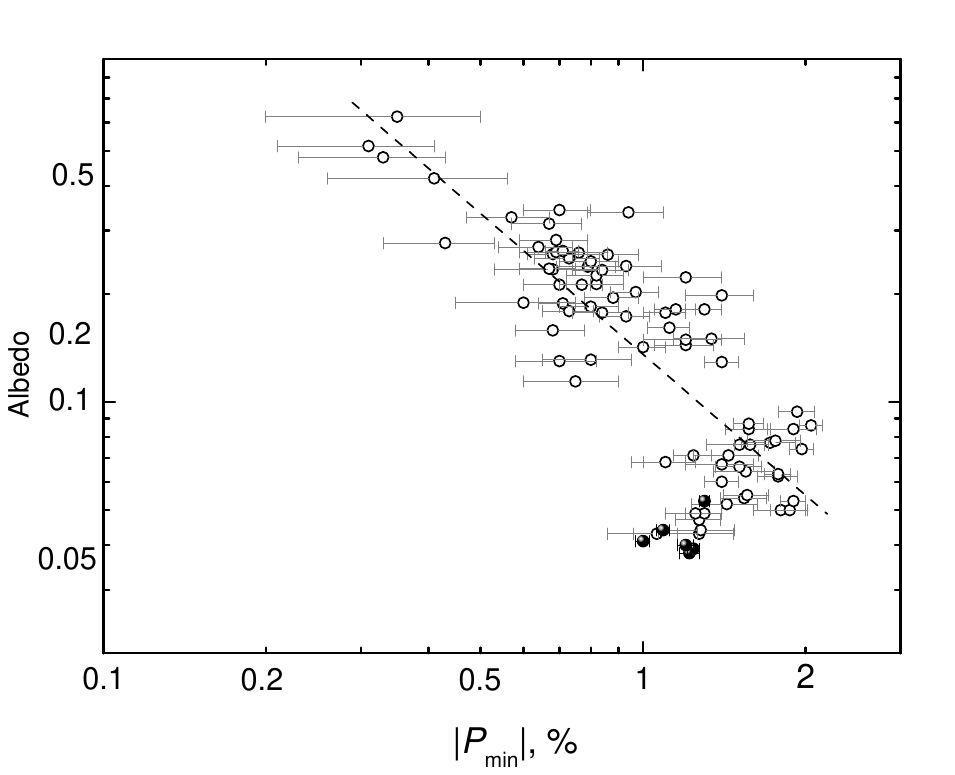}
\caption{\label{Fig_Albedo} 
Relationship between albedo and polarization minimum
$P_{\rm min}$ for asteroids (open symbols) and Trojans (filled symbols).
}
\end{figure}
The relationship of $P_{\rm min}$ and albedo based on the updated data on
albedos is plotted in Fig.~\ref{Fig_Albedo}. Data for Trojans are within the range of
the low albedo asteroids. The saturation effect for low-albedo asteroids is fairly
evident.

If our new estimates of the albedos are correct, then the differences
observed between (4543) Phoinix, (588) Achilles, and the group of
three asteroids (1583) Antilochus, (6545) 1986 TR6, and (21601)
(1998 X089) may just reflect a difference in the surface structure
that could also be revealed, e.g., by a difference in the reflectance
spectra.  The spectral properties of (4543) Phoinix have not been
measured, and (588) Achilles is classified as an unusual D type (DU) by
Tholen (1989). The remaining three asteroids are D type. We therefore
expect that the taxonomic class of (4543) Phoinix may also differ from the
typical D type.



\subsection{Comparison with other atmosphere-less objects in the solar system}
\begin{figure}
\includegraphics*[scale=0.7,trim={0.0cm 0cm 0.0cm 0.0cm},clip]{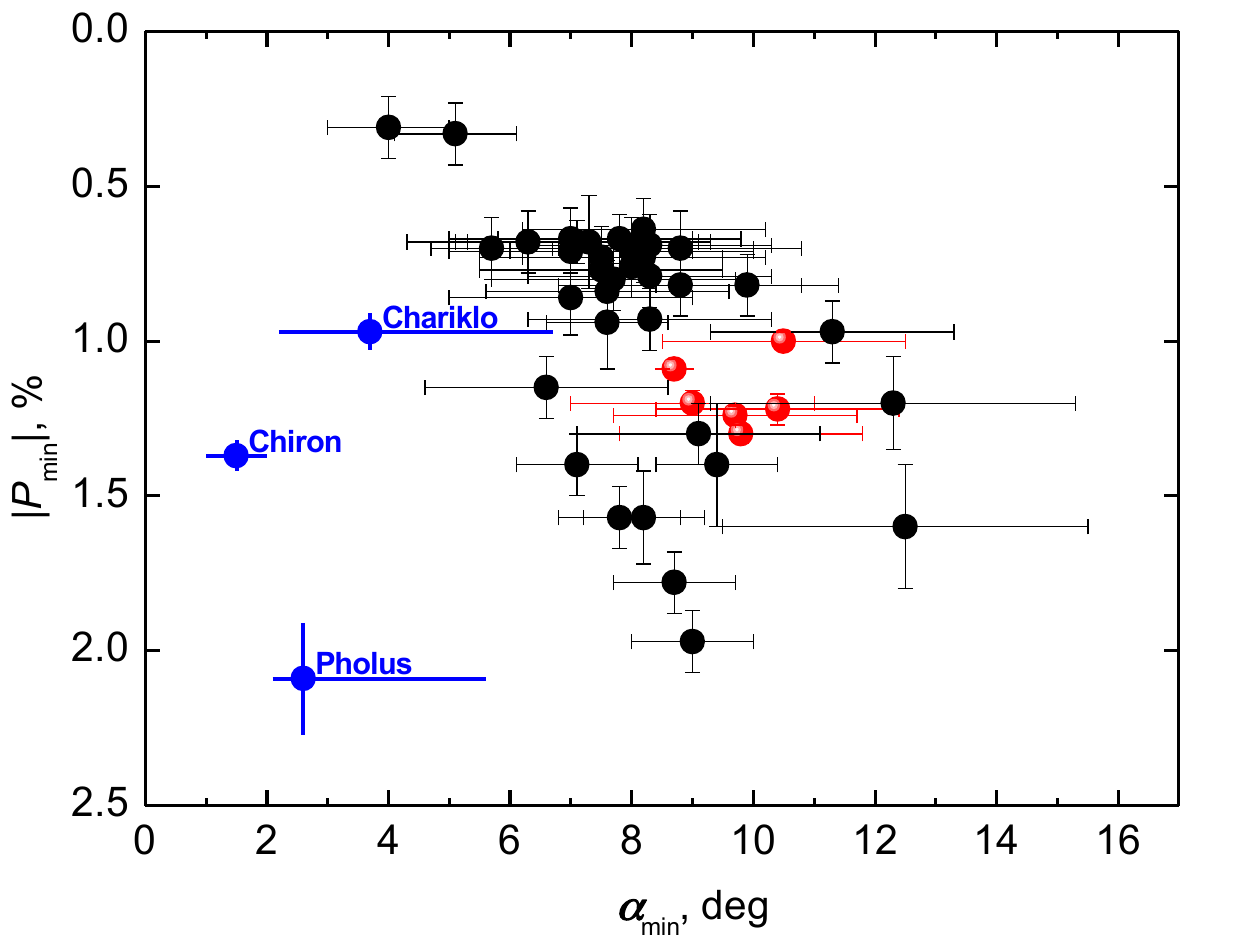}
\caption{\label{Fig_Palfa} 
Depth of the polarization minimum $P_{\rm min}$ versus the phase-angle
$\alpha_{\rm min}$ where the minimum occurs for asteroids (black) , Trojans (red), and Centaurs (blue points).}
\end{figure}

\begin{figure}
\includegraphics*[trim={0.0cm 0cm 0.0cm 0.5cm},clip]{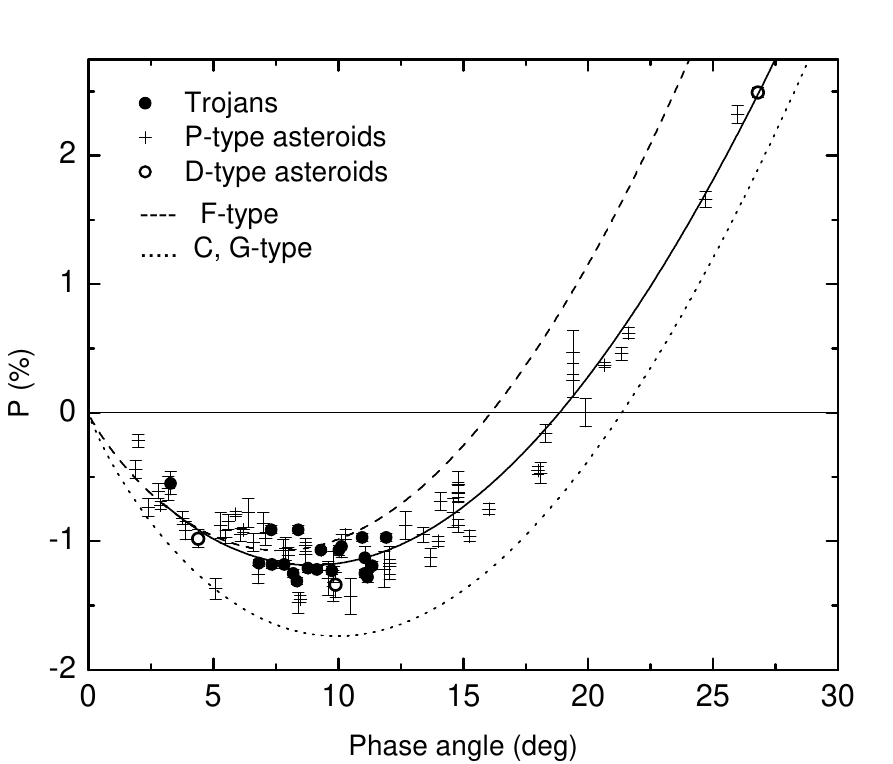}
\caption{\label{Fig_Comparison} Polarization phase angle dependence of
  Jupiter Trojans (filled circles), D-type asteroids (open circles),
  and P-type asteroids (crosses). The fits by a Lumme \& Muinonen
  function are shown for the F-type (dashed line), P-type (solid
  line), and C,G type (dotted line) asteroids.}
\end{figure}
We have compared the polarimetric properties of Trojans to the
literature data on TNOs \citep[][and references therein]{Bagetal08},
Centaurs \citep[][and references therein]{Beletal10}, and low-albedo
asteroids (see database compiled by Lupishko and available at {\tt
  http://sbn.psi.edu/pds/resource/apd.html}).  The mean values of
the polarimetric parameters $P_{\rm min}$, $\alpha_{\rm min}$,
$\alpha_{\rm inv}$ and their scattering are given in Table~\ref{Tab_Compa}.

The polarimetric properties of TNOs and Centaurs are not
characterized as well as those of main-belt asteroids. Because of their
distance, Centaurs can only be observed in a limited phase angle range
($\sim 0-5\degr$, compared to $\sim 0-30\degr$ of main-belt
objects). However, there are indications that the polarimetric curves
of Centaurs reach a minimum at very small phase angles \citep[as small
  as $\sim 1.5\degr$ for Centaur Chiron, see][]{Bagetal06,Beletal10}.
This feature was interpreted by \citet{Beletal10} as indicative of a
thin frost layer of submicron water ice crystals on their dark
surfaces. Both for Trojans and Centaurs, we can only estimate a lower
limit of the inversion angles, since geometrical constraints make
their direct measurements impossible from Earth-based
observations. For TNOs, that with the exception of the binary system
Pluto-Charon are visible from Earth only at phase angles $\la 2$\degr,
we cannot even estimate the parameters $P_{\rm min}$ and $\alpha_{\rm min}$.

Figure~\ref{Fig_Palfa} shows the relationship of polarization minimum
and the phase angle where the minimum occurs for Trojans, Centaurs,
and main-belt asteroids. The polarization-phase angle behaviour of the
observed Trojans is very similar to that of low-albedo asteroids, in
particular the P-type asteroids, and quite different from those Centaurs
for which polarimetric measurements have been obtained, in spite of
closer proximity to the latter group of objects.

\begin{table}
\caption{\label{Tab_Compa} Polarimetric properties of some
  atmosphere-less objects. The number of objects for which
  polarization minima ($P_{\rm min}$ at $\alpha_{\rm min}$) and
  inversion angles ($\alpha_{\rm min}$) were measured is indicated by
  $N_{\rm min}$ and $N_{\rm inv}$, respectively.}
\begin{center}
\begin{tabular}{lrr@{$\pm$}lr@{$\pm$}lcr@{$\pm$}l} 
\hline\hline
&
&
\multicolumn{2}{c}{$P_{\rm min}$}&
\multicolumn{2}{c}{$\alpha_{\rm min}$}&
&
\multicolumn{2}{c}{$\alpha_{\rm inv}$}\\
Object &
$N_{\rm min}$&
\multicolumn{2}{c}{(\%)}&
\multicolumn{2}{c}{$(\degr)$}&
$N_{\rm inv}$&
\multicolumn{2}{c}{$(\degr)$}\\
\hline                       
Trojans        &  5 &$-1.15$&0.15 &9 & 1   &0& \multicolumn{2}{c}{$\ga17$}\\
P-type         & 11 &$-1.29$&0.22 &8 & 2   &4& 19.2& 1.0 \\
F-type)        &  4 &$-1.17$&0.11 &7 & 2   &4& 16.1& 1.4 \\
G- and C-type  & 20 &$-1.70$&0.20 &9 & 2   &9& 20.8& 0.6 \\
Centaurs       &  4 &$-1.44$&0.47 &\multicolumn{2}{c}{$\sim 2$?}
                                           &0&\multicolumn{2}{c}{$\ga6$} \\
Small TNOs &  0 &\multicolumn{2}{c}{$<-1.5$}
                                  &\multicolumn{2}{c}{>2}
                                           &0&\multicolumn{2}{c}{?}\\
\hline
\end{tabular}
\end{center}

\end{table}
Figure~\ref{Fig_Comparison} shows the mean polarization-phase curves
for the P-, F-, G- and C-type asteroids, and demonstrates that the
data for the P-type asteroids and D-type Trojans are practically
indistinguishable. Compared to the F-type asteroids, polarization
minima of Trojans occur at a larger phase-angle, which suggests that their
inversion phase-angles are also larger. \citet{Foretal06} obtained a
polarimetric measurements of the D-type object (944) Hidalgo at a
large phase angle, and (944) Hidalgo has an unusual orbit with a semi-major axis
of 5.74\,au and eccentricity of 0.66. This object reaches 1.94\,au in
perihelion, giving an opportunity to observe it in much larger phase-angle range than for other D types. The polarization measurement at
$ \alpha = 26.8\degr$ lies exactly at the fitted phase curve for
P-type asteroids and confirms a similarity of polarization properties
of D- and P-type asteroids within the accuracy of polarimetric
measurements.

\section{Conclusions}
We performed a pilot study of the polarization properties of
Jupiter Trojan asteroids and have obtained measurements for six objects
belonging to the L4 population. Comparing our targets, we found that they show
similar but not identical polarization properties, in particular that
there are at least two distinct polarimetric behaviours. Trojans
(588) Achilles and (1583) Antilochus show a shallower polarization
curve than the remaining four Trojans (3548) Eurybates, (4543)
Phoinix, (6545) (1998 TR6), and (21601 (1998 X089). The C-type
Trojan (3548) Eurybates shows the deepest minimum of polarization.  D-type
Trojans (1583) Antilochous, (6545) (1998 TR6) and (21601) (1998 X089)
all have a minimum around $-1.2$\,\%, but overall their polarimetric
behaviour does not appear very different from that of (3548)
Eurybates. 
Considering all objects together, we found that the minimum of the
polarization is reached at a phase angle $\sim 10\degr$ and is in the
range of$-1.3$\,\% to $-1.0$\,\%. This polarimetric behaviour is
different from that of Centaurs, which seem to show polarization
minima at much smaller phase angles and are very similar
to low-albedo main-belt asteroids.

\begin{acknowledgements}
Based on observations made with ESO Telescopes at the La Silla-Paranal
Observatory under programme ID 091.C-0687 (PI: I.\ Belskaya).  SB, IB, AS,
and GBB acknowledge support from COST Action MP1104 "Polarization as a
tool to study the Solar System and beyond" through funding granted for
Short Term Scientific Missions and participations to meetings. 

\end{acknowledgements}

\Online
\begin{figure*}[ht]
\begin{center}
\scalebox{0.7}{
\includegraphics*[angle=270,trim={1.5cm 0cm 1.5cm 0cm},clip]{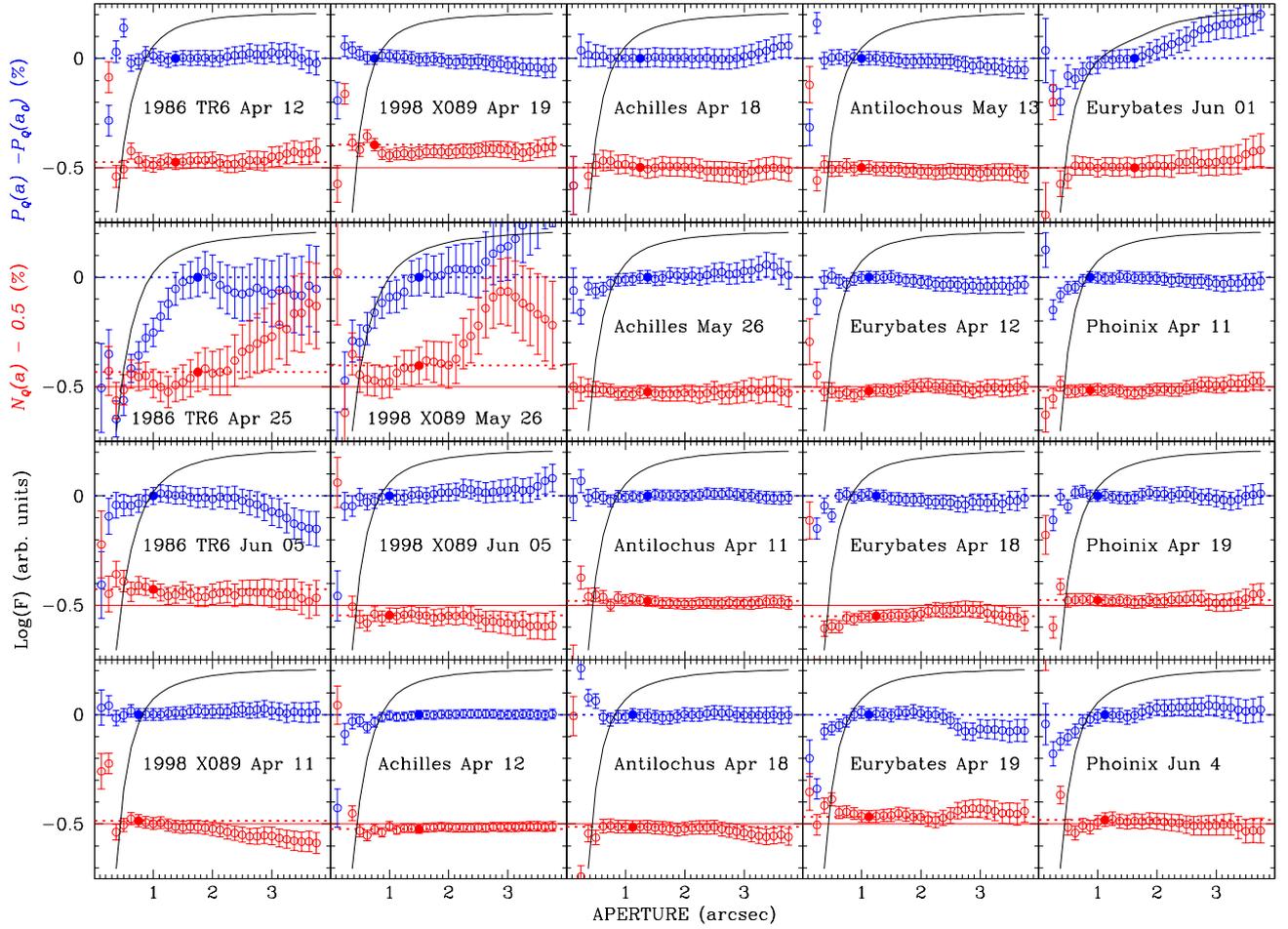}}
\caption{\label{Fig_Apertures_PQ} Aperture polarimetry: \pq\ and \nnq\ parameters as function of the aperture
for the various observing series. \pq\ parameters, represented by blue empty circles,
are offset to the values corresponding
to the aperture adopted for the measurement and reported in Table~\ref{Tab_Observations}. This point is hightlighted with solid circles
and dotted lines. \nnq\ parameters, represented by red empty circles, are offset by $-0.5$\,\%
for display purpose. Again, the adopted values are highlighted with solid symbols and dotted lines.
Each panel of this figure is similar to the left panel of Fig.\ref{Fig_Example_Apertures_QU} and 
is explained in more detail in Sect.~\ref{Sect_Aperture_Polarimetry}.}
\end{center}
\end{figure*}

\begin{figure*}[ht]
\begin{center}
\scalebox{0.7}{
\includegraphics*[angle=270,trim={1.5cm 0cm 1cm 0cm},clip]{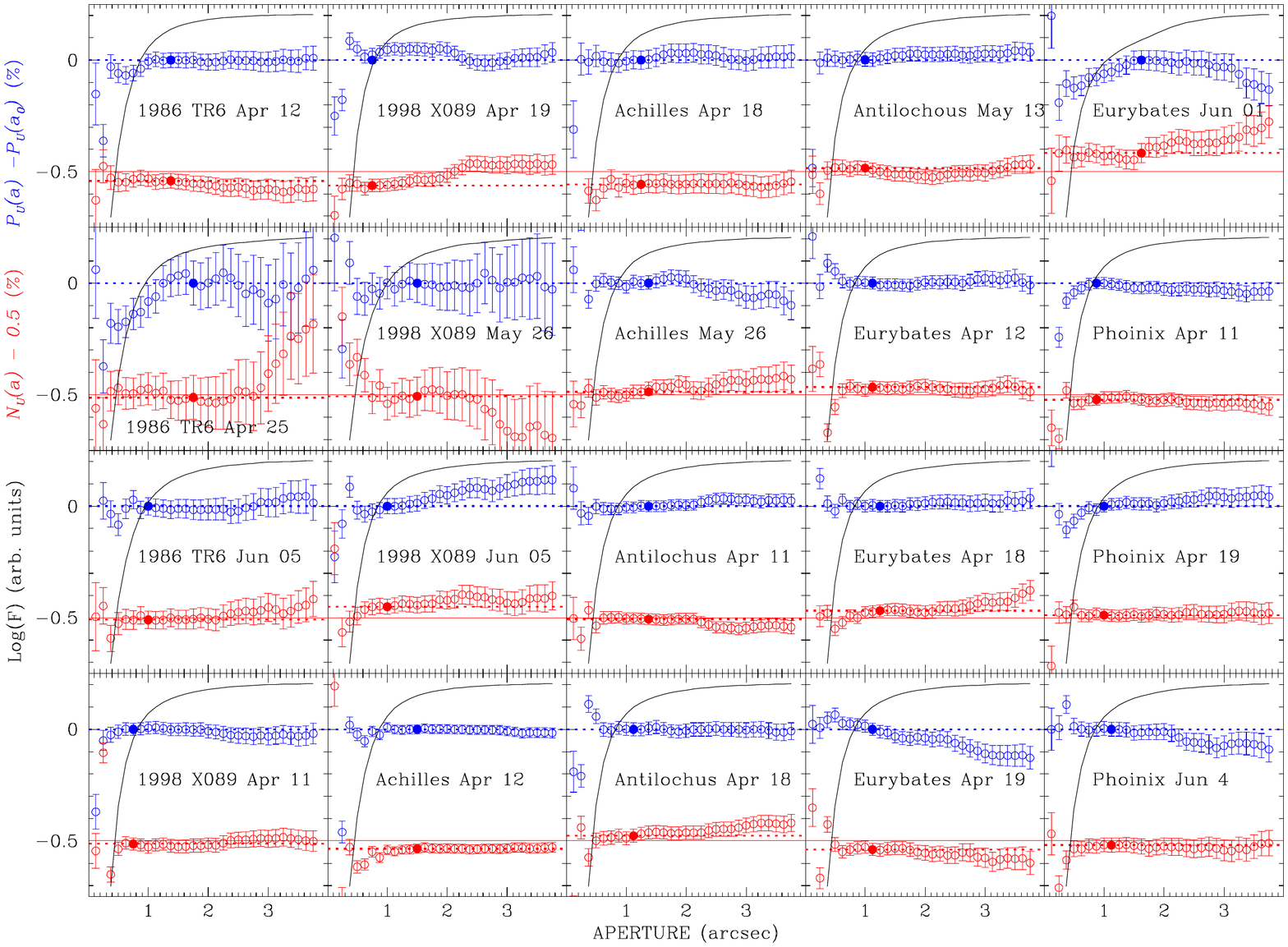}}
\caption{\label{Fig_Apertures_PU}  Aperture polarimetry: \pu\ and \nnu\ parameters as function of the aperture
for the various observing series. \pu\ parameters, represented by blue empty circles, are offset to the value corresponding
to the aperture adopted for the measurement and reported in Table~\ref{Tab_Observations}. This point is hightlighted with solid circles
and dotted lines. \nnq\ parameters are represented by red empty circles, and are offset by $-0.5$\,\%
for display purpose. Again, the adopted values are highlighted with solid symbols and dotted lines.
Each panel of this figure is similar to the right panel of Fig.\ref{Fig_Example_Apertures_QU} and 
is explained in more detail in Sect.~\ref{Sect_Aperture_Polarimetry}.}
\end{center}
\end{figure*}

\end{document}